# Agricultural Growth Diagnostics: Identifying the Binding Constraints and Policy Remedies for Bihar, India*

Elumalai Kannan
Associate Professor, Centre for the Study of Regional Development
School of Social Sciences, Jawaharlal Nehru University (JNU), New Delhi-110067
Email: elumalaik@mail.jnu.ac.in
&
Sanjib Pohit
Professor, National Council of Applied Economic Research (NCAER), New Delhi-110002
Email: spohit@ncaer.org

**Abstract**

Agriculture plays a significant role in economic development of the underdeveloped region. Multiple factors influence the performance of agricultural sector but a few of these have a strong bearing on its growth. We develop a growth diagnostics framework for agricultural sector in Bihar located in eastern India to identify the most binding constraints. Our results show that poor functioning of agricultural markets and low level of crop diversification are the important reasons for lower agricultural growth in Bihar. Rise in the level of instability in the prices of agricultural produces indicates a weak price transmission across the markets even after repealing the agricultural produce market committee act. Poor market linkages and non-functioning producer collectives at village level affect the farmers' motivation for undertaking crop diversification. Our policy suggestions include state provision of basic market infrastructure to attract private investment in agricultural marketing, strengthening the farmer producer organisations, and a comprehensive policy on crop diversification.

**Keywords:** agriculture, growth diagnostics, binding constraints, agricultural markets

**Introduction**

India's agricultural sector has undergone rapid structural transformation over the past six decades. The elements of agricultural transformation included reduction in the contribution of agriculture to national income, a decline in the share of labour and a reduction in rural poverty and malnutrition (Joshi et al., 2004; Gokaran & Gulati, 2006; Otsuka & Yamano, 2006; Hazell, 2009; Birthal et al., 2015). However, changes in the indicators of agricultural transformation are not uniform across the Indian states. Despite being located in the fertile Indo-Gangetic plain, Bihar remains one of the poorest states in India and also scores very low on the composite development index (Government of India, 2013). The average per capita income of Bihar is twenty times less than that of Haryana during 2008-16 (Pohit et al., 2019). Despite a shift in structural composition from primary sector to services sector, agriculture still contributes over a quarter of the state's income and employs about 70 per cent of the rural workforce. A robust growth in agriculture holds the key for the economic and social development of the state (Thakur et al., 2000; Government of India, 2008; Fujita, 2014).

Considering the importance of agriculture in the economy, Government of Bihar has launched multiple development initiatives since 2008 under the so-called agriculture roadmaps[1]. The thrust is on the holistic development of agriculture with an emphasis on increasing productivity growth and improving farmers' income. These initiatives seem to have helped to accelerate the Bihar's agricultural growth (Minato, 2014). Our estimates show that the agriculture sector registered an annual growth of 2.0 per cent during the period from 2000–01 to 2007–08. In the subsequent period from 2008–09 to 2011–12, agricultural growth increased considerably to 3.1 per cent, which led to achieving a very high growth rate of 10.9 per cent in Gross State Domestic Product (GSDP). However, during the subsequent period of five years (2012–13 to 2016–17), agricultural growth decelerated to 1.3 per cent, which also pulled down the overall economic growth to 6.6 per cent.

It is in this context that we attempt to identify the most binding constraints on Bihar's agricultural growth. It is expected that the removal of these binding constraints through policy reforms would unleash the growth potential in the sector (Hausmann & Klinger 2008). We focus on the crop sector as it accounts for over two-thirds of agricultural output and its impact on overall agricultural growth is relatively high as compared to that of the other sub-sectors. This paper makes two important contributions: (i) develop a generic growth diagnostics framework for agricultural sector, and (ii) apply this framework to identify the most binding constraints on Bihar's agriculture and suggest policy remedies.

---

[1] The timeline of agriculture roadmap are as follows: the first- 2008/09 to 2011/12), the second -2012/13 to 2016/17) and the third- 2017/18 to 2022/23. Each of these road maps lay out production targets/milestones through popularisation of various technologies such as quality seeds, machinery, animal breeds and organic inputs through various programmes in a time bound manner.

The rest of the paper is organised as follows. The second section provides the methodological framework, which is a modified growth diagnostics framework of Hausmann et al. (2008a). The third section presents data sources. The fourth section traces the agricultural performance of Bihar with a view to identify the determinants of the crop output growth. The fifth section goes deeper into the diagnostics of agricultural growth. The sixth section discusses the dimensions of most binding constraints. Final section provides policy recommendations.

**Methodological Framework**

Growth diagnostic framework is conceptualised as a decision tree, which follows a top-down approach. This framework starts with determinants of output growth and then identify the distortions that underlie the binding constraints (Hausmann et al., 2008a; Hausmann et al., 2008b). Economic theory and existing empirical evidence help in identification of growth drivers and binding constraints. Although this framework is widely used for analysing the binding constraints affecting overall economic growth of a country (Hausmann & Klinger, 2008; Obuchi et al., 2016; Hausmann et al., 2017), its application for unravelling the binding constraints of a particular economic activity or sector at a regional level is much limited. While developing a growth diagnostics framework for agriculture, we retain the key elements of the Hausmann et al. (2008a) framework.

The starting point of our modified growth diagnostics framework is the Minot et al. (2006) approach. This approach helps to analyse the drivers of output growth and then diagnose which of these forces pose the greatest obstacles to higher growth. The next step is to uncover the distortions associated with these growth constraints with the assumption that removal of these distortions would unleash growth. As the Minot et al strategy focuses only on the gross revenue side of growth analysis, it can easily be extended to incorporate the cost side as well.

A hybrid of the Minot and Hausmann framework for diagnosing the agricultural growth is depicted in Figure 1. Application of this framework involves asking a series of questions about the binding constraints on growth determinants. Suppose that the overarching problem in Bihar is why agricultural growth has slowed down in recent years? If the problem seems to be the low scale of farming (land constraint), is that due to low cropping intensity, which in turn is due to either poor soil quality, inadequate irrigation facility, expensive labour or government restriction on a particular cropping pattern? Is low scale of farming also due to insecure land tenure, fragmented landholdings, high rent or restrictions on land leasing?

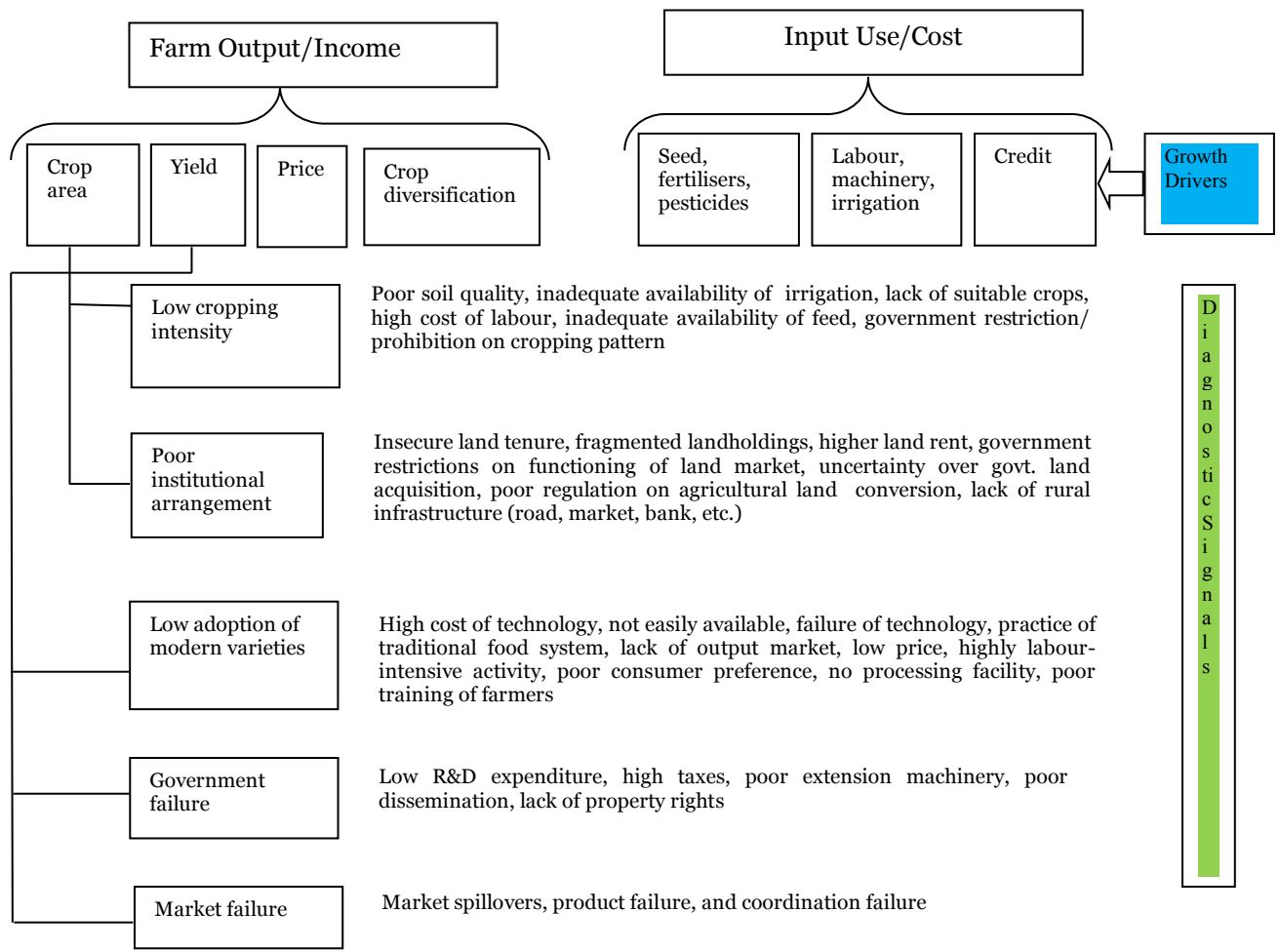

**Figure 1.** Minot and Hausmann hybrid framework for agricultural growth diagnostics

**Source:** Developed by authors

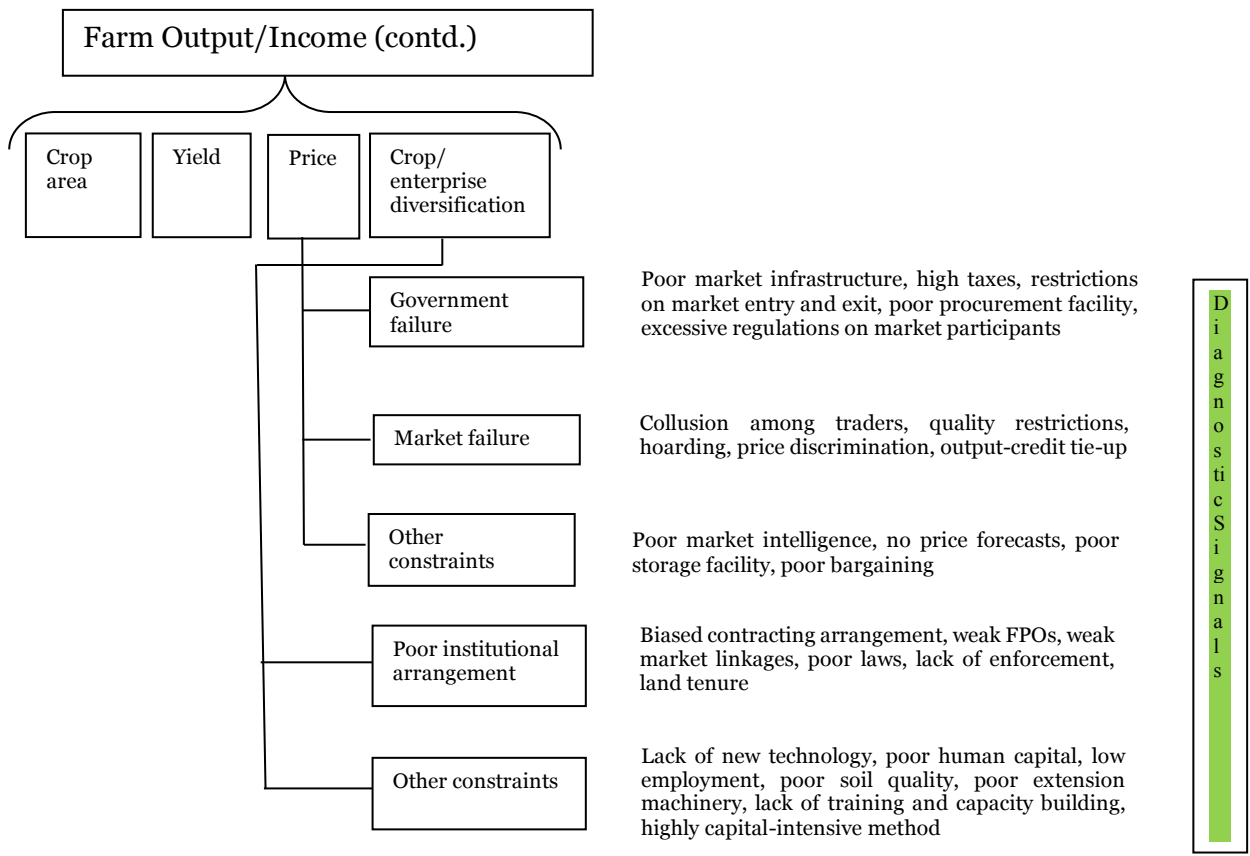

**Figure 1.** Minot and Hausmann hybrid framework for agricultural growth diagnostics *(Continued)*

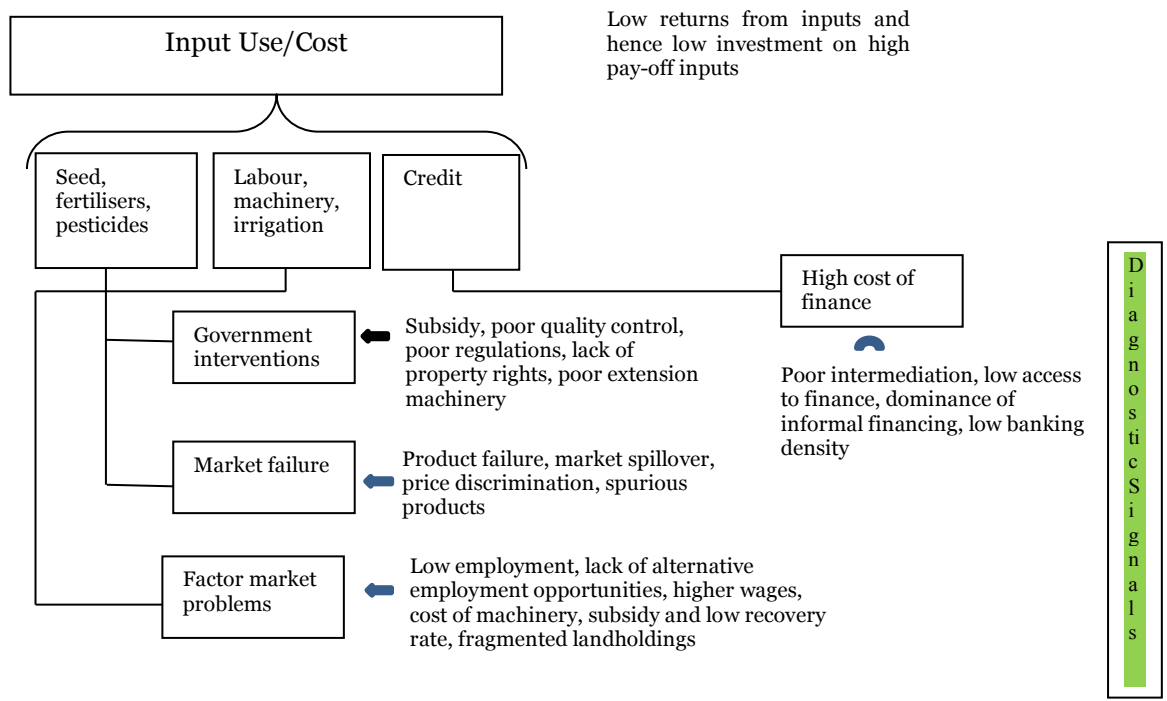

**Figure 1.** Minot and Hausmann hybrid framework for agricultural growth diagnostics *(Continued)*

If low crop yield appears to be a problem, is that due to lack of access to new technology, high cost of technology, low agricultural research and development expenditure, high taxes or poor definition of property rights? Similarly, binding constraints on other determinants of output growth can be identified. The knowledge thus, generated can be used to locate remedies to overcome the obstacles to growth.

On the input side, farmers could face a situation where they get lower returns from inputs use and hence low motivation, resulting in underinvestment on high pay-off inputs. If the problem is with non-availability of quality inputs, is that due to poor quality control, corruption, poor delivery system or high cost? Similarly, if the high cost of financing is a problem, is that due to poor intermediation, low banking density or the dominance of informal financing? In a similar way, binding constraints on specific inputs such as fertilisers, seed, labour, irrigation and machinery can be identified.

The broad framework discussed above is complemented with the following analytical tools. A growth accounting approach proposed by Minot et al. (2006) is followed to analyse the sources of crop output growth. According to this approach, change in gross crop income (dR) from *n* crops can be decomposed as follows.

$$dR \cong \left(\sum_{i=1}^{n} a_i Y_i P_i\right) d\left(\sum_{i=1}^{n} A_i\right) + \sum_{i=1}^{n} A_i \sum_{i=1}^{n}(a_i Y_i dP_i) + \sum_{i=1}^{n} A_i \sum_{i=1}^{n}(a_i P_i dY_i) + \sum_{i=1}^{n} A_i \sum_{i=1}^{n}(Y_i P_i da_i) \quad (1)$$

Where, $A_i$ is the area under crop *i*, $Y_i$ is production per unit area, $P_i$ is the real price and $a_i$ is the share of crop *i* in the total cropped area, i.e., $\left(\frac{A_i}{\sum_i A_i}\right)$.

The first term on the right-hand side of equation (1) represents change in the total cropped area, while the second term measures effect of change in the real prices of commodities on gross revenue. The third term measures changes in the crop yields or technology. Finally, fourth term represents change in the crop composition, implying a re-allocation of land between crops.

Further, total factor productivity (TFP) growth was estimated for major crops by using the Tornqvist-Theil index to analyse the contribution of technological change to output growth. The logarithmic form of the index is given as follows:

$$\ln\left(\frac{TFP_t}{TFP_{t-1}}\right) = \sum_j R_j \ln\left(\frac{Y_{jt}}{Y_{jt-1}}\right) - \sum_i S_i \ln\left(\frac{X_{it}}{X_{it-1}}\right) \quad (2)$$

where, $R_j$ is revenue share of j$^{th}$ output, $S_i$ is cost share of i$^{th}$ input, $Y_{jt}$ is output and $X_{it}$ is input measured, all in period $t$.

Here, total output growth is estimated by summing the growth of each output weighted by its revenue share while the input growth is estimated by summing the growth of each input weighted by the cost share.

**Data Sources**

The present study relies on both secondary and primary data. Data on crop production, value of output and farm harvest prices for 31 crops were compiled from the Directorate of Economics and Statistics, Ministry of Agriculture and Farmers' Welfare, and Central Statistics Office, Government of India. Detailed information on inputs and output for six major crops, viz., paddy, wheat, maize, gram, lentil and potato were collated from the cost of cultivation survey. To ensure the consistency in our analysis, we use data for the post-state bifurcation period from 2000–01 to 2016–17.

A field work[2] comprising in-depth discussions with various stakeholders was conducted during February-July 2019. Details of these stakeholder discussions include: 24 village level focus group discussion (FGD) with farmers; 24 personal interviews with experts in the sectors of seed, fertiliser, irrigation, marketing, processing, horticulture, agricultural technology and credit. The field work was conducted primarily for exploring various viewpoints and ground-level experiences in identifying the binding constraints and possible options for removing these constraints. The results from the analysis of qualitative information were useful to explain the findings from the quantitative analysis.

**Relative Performance of Agriculture**

Agriculture holds the key to the overall development of the Bihar economy. The average annual growth in agriculture and allied activities during the pre-agriculture road map was only about 2.0 per cent (Table 1). During the period of the first agriculture road map, the growth rate accelerated to 3.1 per cent, which was almost equal to the national average agricultural growth. However, this higher growth in agriculture did not sustain in the long run. The average annual growth rate declined to 1.28 per cent during the period of the second agriculture road map. During the overall period 2001-02 to 2016-17, average growth was only 2.0 per cent, which was much below the national average agricultural growth of 3.1 per cent. The moot question is what explains the fall in agricultural growth despite a stable political environment, improvement in investment on rural infrastructure, and reforms in agricultural marketing (Intodia, 2012; Fujita, 2014).

---

[2] Field work was carried out in eight districts spread in all the four agro-climatic zones (ACZ). They include West Champaran and Samastipur (ACZ-I), Purnea and Khagaria (ACZ-II), Bhagalpur and Jamui (ACZ-IIIA), and Bhojpur and Patna (ACZ-IIIB). Three villages from each district were selected.

**Table 1.** Average annual growth in major sectors of the Bihar economy

| Sectors | Agricultural road map period | | | Overall 2001/02 – 16/17 |
|---|---|---|---|---|
| | Pre (2001/02 – 07/08) | First (2008/09 – 11/12) | Second (2012/13 – 16/17) | |
| *Agriculture & allied* | 1.98 | 3.11 | 1.28 | 2.04 |
| Industry | 8.78 | 14.18 | 6.09 | 9.29 |
| Services | 6.38 | 14.56 | 7.65 | 8.82 |
| *Non-agriculture* | 6.93 | 14.44 | 7.02 | 8.83 |
| *Overall* | 4.68 | 10.86 | 6.56 | 6.81 |

**Source:** National Accounts Statistics, Central Statistics Office, Government of India

*Changes in Relative Share of crop area*

Three crops viz., paddy, wheat and maize account for about 70 per cent of total cropped area and 40 per cent of total value of crop output (Table 2). The importance of paddy among the farmers has come down marginally. The area under paddy is being shifted to the cultivation of maize. Farmers increasingly prefer to grow maize due to its growing demand in the food processing industry and as poultry feed (Hellin and Erenstein, 2009). Similarly, the area under wheat has increased and it constituted over a quarter of the total cropped area. The decline in area under rabi (winter) coarse cereals has been compensated by a rise in wheat area.

Between 2002-03 and 2016-17, the area under pulses has declined by over 20 per cent despite a significant rise in their minimum support prices in the past few years (Government of India, 2016). Similarly, the area under jute, cultivated in north Bihar, has declined considerably. Lack of a proper policy and institutional support, and weak markets are responsible for the decline in jute cultivation (Sarkar, 1986; Jha & Viswanathan, 1999).

**Table 2.** Relative share of crop area and value of output

| Particulars | % share of crop area | | | % share of value of output | | |
|---|---|---|---|---|---|---|
| | TE 2002-03 | TE 2007-08 | TE 2016-17 | TE 2002-03 | TE 2007-08 | TE 2016-17 |
| Paddy | 45.3 | 44.5 | 43.0 | 20.4 | 19.7 | 20.8 |
| Wheat | 26.5 | 27.2 | 27.8 | 13.7 | 16.1 | 13.0 |
| Maize | 7.6 | 8.4 | 9.3 | 3.6 | 4.5 | 5.9 |
| Total Cereals | 80.1 | 80.7 | 80.4 | 37.9 | 40.3 | 39.9 |
| Moong | 2.4 | 2.3 | 2.2 | 1.1 | 0.8 | 1.1 |
| Lentil | 2.2 | 2.1 | 2.0 | 1.2 | 1.1 | 1.5 |
| *Khesari* | 1.9 | 1.3 | 0.8 | 0.4 | 0.3 | 0.3 |
| Total Pulses | 8.8 | 7.9 | 6.8 | 4.3 | 3.6 | 3.9 |
| Total Food grains | 88.9 | 88.6 | 87.3 | 42.2 | 43.9 | 43.8 |
| Jute | 1.8 | 1.7 | 1.2 | 0.8 | 1.1 | 1.2 |
| Total Fibres | 2.2 | 2.0 | 1.5 | 0.9 | 1.2 | 1.4 |
| Rapeseed & mustard | 1.2 | 1.1 | 1.1 | 0.6 | 0.7 | 0.8 |
| Total Oilseeds | 1.8 | 1.8 | 1.5 | 1.0 | 1.2 | 1.0 |
| Sugarcane & Gur | 1.3 | 1.5 | 3.2 | 2.1 | 1.6 | 3.4 |
| Potato | 1.8 | 1.9 | 4.2 | 2.1 | 2.7 | 2.4 |
| Fruits & Vegetables | 5.1 | 5.4 | 6.0 | 47.5 | 42.3 | 42.0 |
| Horticulture | 5.2 | 5.6 | 6.2 | 47.7 | 42.5 | 42.3 |
| Others | 0.3 | 0.3 | 0.2 | 7.0 | 10.8 | 9.5 |
| Overall | 100.0 | 100.0 | 100.0 | 100.0 | 100.0 | 100.0 |

**Source:** Computed based on Government of India (various years)
**Note:** TE-Triennium Ending

However, the area under sugarcane has increased considerably during recent years. The share of sugarcane in total cropped area has risen from 1.3 per cent to 3.2 per cent between 2002-03 and 2016-17. Availability of groundwater and an increase in the number of sugar mills are partly responsible for the increase in area under sugarcane (Solomon, 2016). With favourable climatic conditions and natural resources, Bihar is highly suitable for the cultivation of fruits and vegetables. Although the area under fruits and vegetables constituted about 6.0 per cent of total cropped area, they contributed over 40 per cent of total value of output. However, level of crop diversification towards cultivation of fruits and vegetables is very low.

*Sources of output growth*

Sources of crop output growth are provided in Table 3. Between 2001-02 and 2007-08, contribution of price to crop output growth was significantly higher along with diversification. These effects could not help much in boosting output growth due to large negative effect of yield in this period. However, output growth of about 3.6 per cent in 2008-09 to 2016-17 has been bolstered with the impressive contribution of yield effect. The contribution of diversification effect has also improved and has positively influenced output growth. Improvement in diversification effect shows the extent of area reallocation by farmers from low-productive crops to high-productive crops such as horticultural crops. The contribution of real price and area expansion to overall output growth was negative in the recent period implying they cease to be a positive factor for growth.

**Table 3.** Sources of crop output growth

| Particulars | 2001-02 to 2007-08 | 2008-09 to 2016-17 | 2001-02 to 2016-17 |
|---|---|---|---|
| Area effect | -6.9 | -7.8 | -7.5 |
| Diversification effect | 8.1 | 36.8 | 26.9 |
| Yield effect | -54.4 | 210.8 | 119.3 |
| Price effect | 176.1 | -77.2 | 10.2 |
| Interaction effect | -23.0 | -62.6 | -48.9 |
| Total | 100.0 | 100.0 | 100.0 |

**Source:** Authors' estimates

On the whole, yield effect was dominant, along with positive diversification and price effects. The contribution of price effect to output growth is lower at 10.2 per cent than that of yield effect. The negative interaction effect is largely due to a fall in the contribution of area effect to output growth. Diversion of productive agricultural land for non-agricultural uses and increase in fallow land are responsible for a fall in cultivated land (van Duijne, 2019).

**Diagnostics of Agricultural Growth**

Having examined the sources of crop output growth, we now focus on distortions associated with these growth drivers through the lens of the diagnostics framework. The question is whether any or multiple of these factors explain the slow-down of agricultural growth in Bihar.

To begin with, can the poor performance of agriculture be explained by decline in agricultural land? Similar to other Indian states, there is increased competition for land between agricultural and non-agricultural uses in Bihar as well (Hoda et al., 2017; van Duijne, 2019). However, the proportion of non-agricultural land to total land area remained constant in Bihar (Table 4). The land use pattern in Bihar is not dissimilar to the trend observed at the national level.

**Table 4.** Ratio of agricultural land and non-agricultural land in total reported area

| Year | Bihar | | India | |
|---|---|---|---|---|
| | AL/Total Area | NAL/Total Area | AL/Total Area | NAL/Total Area |
| TE 2002-03 | 0.68 | 0.18 | 0.55 | 0.08 |
| TE 2007-08 | 0.68 | 0.18 | 0.54 | 0.08 |
| TE 2014-15 | 0.67 | 0.18 | 0.54 | 0.09 |

**Source:** Computed based on Government of India (various years)
**Note:** AL-Agricultural land; NAL-Non-agricultural land; Agricultural land includes net sown area and fallow land

Analysis has shown that overall effect of crop area on crop output growth was negative. Further, there was significant variation in the relative shift of area across the crops. With near stagnation in net sown area, the area gains for wheat, maize, sugarcane and vegetables took place through reallocation of existing amount of land by the farmers. There is no scope for further expansion of cultivated land. However, one important problem that Bihar agriculture faces is fragmented landholdings. This can be a potential problem leading to low economies of scale, but this cannot be a source of low agricultural growth as small sized farms are found to be more productive (Chand et al., 2011; Dev, 2012). Therefore, it is clear that agricultural land is unlikely to be a causal factor for the slow-down of agricultural growth in Bihar.

Can low yield of major crops be the reason for a lower agricultural growth? If this is the case, one would expect low level of technological innovation and limited use of material inputs. Technological change in agriculture is generally analysed through changes in total factor productivity (TFP). A lower TFP growth, keeping the effects of material inputs constant, would indicate lower yield growth and consequently a lower output growth (Fuglie, 2019).

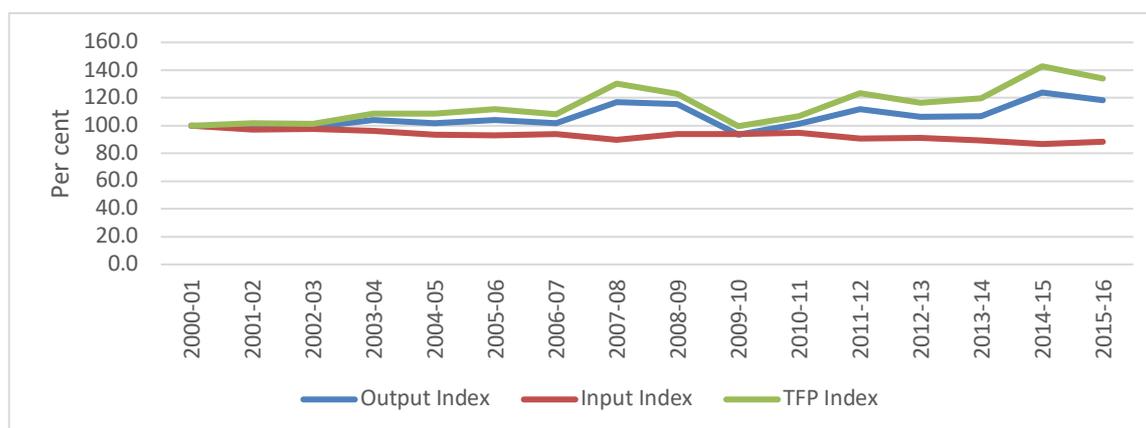

**Figure 2.** Trend in output, input and TFP Index

**Source:** Computed by authors

The trend in weighted indices of output, input and TFP is presented in Figure 2. The aggregate output index has shown a gradual rising trend from 2000-01 to 2005-06. Output index has increased in subsequent years and then declined in 2009-10, largely due to widespread drought in different regions of Bihar. Encouragingly, the output index surged upward thereafter, which corresponds with the second agriculture road map. In case of aggregate input index, it has declined steadily during the entire period. This implies that output growth is largely driven by technological change and that the contribution of input intensification is very low.

The average annual growth in TFP was 1.71 per cent between 2000-01 and 2015-16. This growth is very much comparable with all India agricultural TFP growth of 1.60 per cent (Fuglie, 2019). This shows that Bihar agriculture tends to catch up with technological progress at the national level. Therefore, it is clear that slow or lack of technological progress is not the basic reason for slow-down of agricultural growth.

India's agricultural markets have long been regulated by the government-regulated agricultural produce market committees (APMCs). These markets come under the purview of state governments, which formulate policies to ensure fair trading, transparency in transactions and provision of the necessary infrastructure facilities leading to better price discovery. However, multitude of problems have affected the effective functioning of the APMCs. Some of the well-documented problems of APMCs included collusion of traders, malpractices in transactions, high market fees, poor infrastructure, and diversion of market fees for development works other than markets and lack of competition (Banerji and Meenakshi, 2002; Chand, 2003; Acharya, 2004).

Many reforms were introduced to promote competition in agricultural markets and ensure better price for farmers through legislative measures. For instance, the Model Act on the State Agricultural Produce Marketing (Development & Regulation Act 2003) contained far-reaching reforms, among others, to provide a level playing field for farmers, rationalise the structure of market fees and encourage private investment for creating the necessary infrastructure. While most state governments amended their Acts to incorporate these suggested measures, the government of Bihar was the only state to repeal the APMC Act itself in 2006.

Did these reforms improve price efficiency in the agricultural markets of Bihar? With the abolition of the APMC Act, one would expect that grain markets in Bihar are integrated within the state and also with markets in other states. Farmers are free to sell to traders in any part of Bihar and elsewhere in the country. This would imply that there is an effective price transmission between the grain markets within the state and hence better price received by the farmers. Further, with integration of markets, volatility in grain prices will be low through better flow of information about the supply and demand conditions across the markets.

**Table 5.** Average wholesale price before and after repeal of the APMC Act

| Commodity | Before Repeal (2002-06) | | After Repeal (2007-16) | |
| --- | --- | --- | --- | --- |
| | Average Price (Rs/ton) | Coefficient of Variation (%) | Average Price (Rs/ton) | Coefficient of Variation (%) |
| Paddy | 511 | 11.0 | 1154 | 27.7 |
| Wheat | 771 | 12.2 | 1279 | 14.1 |
| Maize | 600 | 11.2 | 1084 | 24.9 |

**Source:** Computed from agmarknet.gov.in, Government of India

The average price[3] of major crops such as paddy, wheat and maize has increased in the post-market reforms period as compared to the pre-reform period (Table 5). However, volatility in grain prices has also increased, which is evident from the increase in the value of coefficient of variation. Instability in prices of agricultural produce affects the farmers' decision to allocate area under different crops and adopt improved cultivation practices (Acharya, 2004; Acharya et al., 2012; Kishore, 2004). Therefore, instability in the prices of agricultural commodities could be a reason for the lower agricultural growth in Bihar.

Crop diversification is another important element contributing to output growth. The farmers' decision to diversify from cereals to horticultural crops is largely determined by relative profitability of crops and availability of secured marketing arrangements (Joshi et al., 2006; Birthal et al., 2015). Can a low level of crop diversification be the reason for low crop output growth in Bihar? Analysis has shown that crop diversification contributes over a quarter of crop output growth. However, looking at the relative share of crop area gives an impression that the level of crop diversification is very low. In fact, three crops, viz., paddy, wheat and maize, dominated the cropping pattern (80 per cent of cropped area) during 2016-17.

---

[3] Consistent data on wholesale prices of different agricultural commodities are not available after the abolition of APMCs. We could compile marketwise data only for paddy, wheat and maize from the government portal agmarknet.gov.in.

Is there a scope for the state of Bihar to go for greater diversification towards high-value horticultural crops? To examine this, an index of comparative advantage in growing horticultural crops has been developed. The comparative advantage index (CAI)[4] is defined as the ratio of area share of a crop in Bihar to area share of the crop in the country as a whole. The total horticultural area was used as the base value for working out the share. An index value of greater than one for a particular crop/group indicates comparative advantage in growing of that crop.

**Table 6.** Comparative advantage in growing horticultural crops in Bihar

| Particulars | 2013-14 | 2014-15 | 2015-16 |
| --- | --- | --- | --- |
| Fruits | 0.88 | 0.98 | 1.01 |
| Vegetables | 1.82 | 1.76 | 1.72 |
| Flowers | 0.07 | 0.11 | 0.08 |
| Aromatic & medicinal plants | 0.20 | 0.12 | 0.01 |
| Spices | 0.09 | 0.08 | 0.10 |

**Source:** Computed based on Government of India (various years)

It can be seen from Table 6 that Bihar has very high relative advantage in growing vegetables. But the value of the CAI for vegetables has declined from 1.82 in 2013-14 to 1.72 in 2015-16. There is some improvement in relative advantage in the cultivation of fruits during the recent year. Even though Bihar has rich alluvial soil, groundwater, favourable climatic conditions, and is well connected to national capital, the relative importance of horticultural crops is found to be low. Therefore, a low level of crop diversification can be another reason for a low agricultural growth in Bihar.

An important feature of our growth diagnostic framework is that it considers the input use pattern as well. The agriculture road maps focus on, among others, the distribution of seed, fertilisers, pesticides, machinery and provision of credit to farmers. Evidence shows that use of these inputs and mechanisation of agricultural operations has increased overtime (Biggs et al., 2011; Reddy et al., 2014) leading to a rise in the cost of cultivation.

Can rising input costs be a reason for the lower output growth in Bihar? Increase in cost of inputs leads to reduction in profitability. Low profitability affects the decision of the farmers to invest in productivity-enhancing inputs such as irrigation, improved seeds and fertilisers.

---

[4] $CAI_{ij} = (X_{ij}/X_j) / (X_{iw}/X_w)$, where, $X_{ij}$ is area under crop j in the state i, $X_i$ is total crop area of state i, $X_{iw}$ area of crop j in the country and $X_w$ is total crop area in the country.

Low farm investment, further leads to reduction in both the quantity as well as quality of crop output. This also in a way affects the farmers' motive to diversify the cropping pattern and adopt new technological practices.

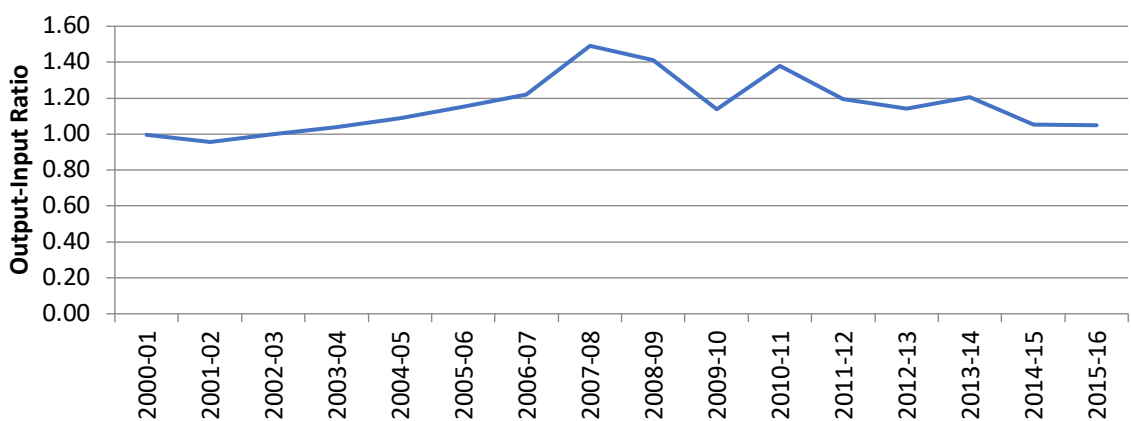

**Figure 3.** Ratio of gross value of output to total input costs

**Source:** Computed based on CACP data, Ministry of Agriculture and Farmers Welfare, Government of India

To analyse the extent of increase in input costs, we take the ratio of value of aggregate crop output to aggregate input costs. This ratio showed an increasing trend until 2007-08 and thereafter it started fluctuating with a declining trend (Figure 3). It has, however, remained above 1, indicating that proportionate increase in output value is higher than total inputs, though it has weakened during recent years. This implies that profitability in crop cultivation has declined.

The use of purchased inputs in the cultivation of crops has increased over time (Sen & Bhatia, 2004; Kannan, 2015; Government of India, 2017). Since it is difficult to analyse the price of all the individual inputs, ratio of fertiliser price[5] (urea) to grain price is analysed here to know if input costs affect the growth performance. It can be seen from Figure 4 that the ratio of wholesale price of wheat to urea price was more than 1 throughout the study period. A similar trend could also be observed in the relation of paddy price to urea price. This indicates that grain price was higher than the fertiliser price. This corroborates the earlier finding that output value has proportionately risen more than the input costs. Further, analysis of drivers of output growth presented in the earlier section has clearly shown that input intensification is low. These findings show that rising input costs cannot be the reason for lower agricultural growth in Bihar.

---

[5] Among various type of fertilisers type, application of urea is relatively high among the farmers (Sharma & Thaker, 2010; Chand & Pavithra, 2015).

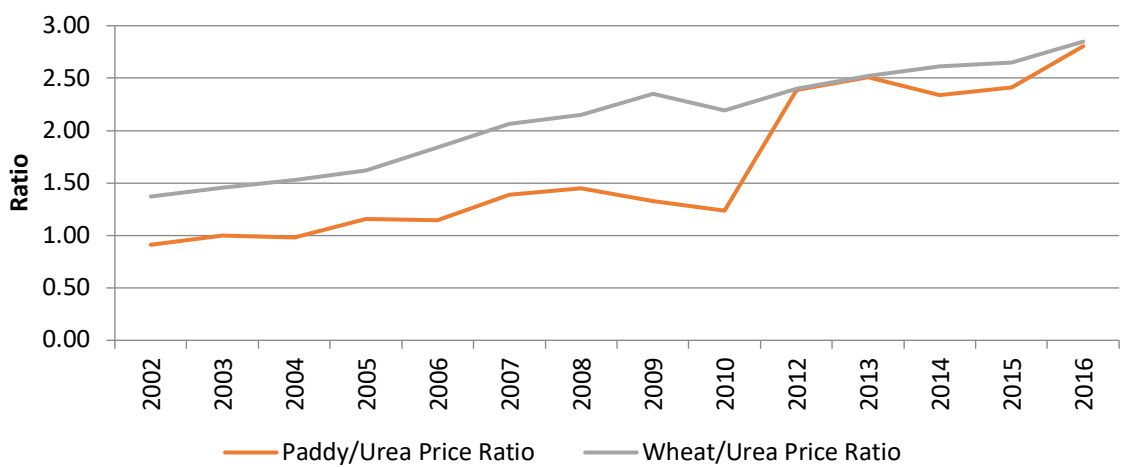

**Figure 4.** Grain to fertiliser price ratio

**Source:** Computed based on CACP data, Ministry of Agriculture and Farmers' Welfare, Government of India

It is clear from the analysis that poor functioning of agricultural markets gleaned through instability in the prices of agricultural products and low level of crop diversification are the reasons for slow or lower agricultural growth in Bihar. Next, it is important to explain why Bihar is constrained in agricultural markets and crop diversification. Relaxing the constraints on agricultural markets and the drive towards crop diversification is expected to result in higher growth in agriculture.

**Explaining the Binding Constraints**

This section analyses the distortions associated with the most binding constraints on Bihar agricultural growth.

*Constraints on Agricultural Markets*

The repeal of APMC Act in 2006 did not usher private investment to create new markets and/or to strengthen facilities in the existing markets. Focus group discussions with farmers and traders revealed that agricultural markets are located far away from the villages and that a particular market serve a large number of villages. Reportedly, there were no storage facilities available in the surveyed villages. Even though limited private warehouse facilities were available in a radius of about 25-30 km from these villages, the storage cost[6] in private warehouses is too high to afford for small farmers. Only, large landholders and traders could avail of the warehouse facility.

Over 90 per cent of the harvests of paddy, wheat, maize, lentil, gram, mustard and banana is sold within the village to traders and commission agents. Most farmers reported that their poor economic conditions and the need for immediate cash after harvest compel them to sell at a lower price to these market agents. Further, government market facilities are not available near the village. Even if farmers take their produce to a distant market yard, they need to bribe the commission agents for facilitating the transactions. Farmers also cannot store produce at their household due to lack of space and poor storage conditions. Therefore, they are forced to sell at whatever the price the traders are willing to offer.

Further, participation of government agencies in procurement of grains and the scale of procurement are reportedly very low. In Bihar, Primary Agriculture Cooperative Societies (PACS) are entrusted with procurement operation, particularly of paddy and wheat from the farmers at the government announced minimum support price (MSP). Ground-level evidence shows that the procurement operation is limited to some fixed quantity and specific time period; these restrictions are considered to be highly arbitrary and vary across villages. The PACS do not procure wheat, which otherwise should do, when there is a glut in the market. This results in price crash. Further, even at PACS, farmers reportedly received a price much lower than the MSP and payments are not made in time. An important function of PACS is to provide crop loan to farmers. In case farmers have taken loan, the final settlement of sale proceeds is made after deducting the loan amount. These issues potentially deter the farmers from selling their harvests to PACS.

Under these situations, farmers are left to the mercy of traders who unscrupulously fix a lower price for agricultural produce that they buy from them. Farmers mentioned that non-availability of a fair price is the most important constraint in expanding agricultural output. Inadequate market facilities and poor institutional arrangements for procurement are responsible for low price realisation and instability in prices.

---

[6] Monthly storage cost for potato ranged from Rs. 160/quintal to Rs. 280/quintal in different survey districts.

*Constraints on Crop Diversification*

Although crop diversification acts as a cushion against unforeseen climatic events, it also entails investment in new technology and marketing arrangements (Rahman, 2009). Evidence shows that crop diversification has potential to increase farm income and reduce poverty (Joshi et al., 2004; Birthal et al., 2015). So, the question is what constraints the farmers to keep crop diversification presently at a low level. One of the nodes shown in the growth diagnostics framework under crop diversification pertains to poor institutional arrangements, particularly weak market linkages and ineffective producer organisations. The Government of Bihar has launched initiatives to establish Farmer Producer Organisations (FPOs) in different parts of the state since 2011-12. FPOs enable farmers to innovate, diversify and adopt new agricultural practices to produce better quality products and reap the benefit of economies of scale (Prasad & Prateek, 2019; Govil et al., 2020). FPO is a potential medium to diversify crop production activities as the marketing activities are collectively taken care of. So, what are the constraints for efficient functioning of FPOs?

Field evidence shows that farmers were aware of FPOs in some villages, but they have not come together to constitute an FPO. Out of 24 villages, FPOs were registered only in 6 villages; none of which were found to be functional. Most farmers were optimistic that FPOs can play a role in reducing the current problems in marketing of agricultural produce. But lack of such organisational set-up on the ground is a constraint on collective bargaining to get a better price. Since traditional crops such as rice, wheat and maize have, by and large, secured markets, area diversion for growing of new crops comes with some risks. This is particularly true in case of vegetables whose prices fluctuate often due to demand and supply gaps. So, lack of collective marketing through FPOs demotivates farmers from going in for a profitable crop diversification.

Similarly, farmers in 10 villages mentioned that they were aware of contract farming. But the practice of contract farming was not reported in any of the surveyed villages. There are no proper policy and suitable legislative measures put in place to promote contracting arrangements in the state. These could be the probable reasons for why agro-business firms not showing an interest in contract farming in Bihar. Most farmers in the surveyed villages said that contract farming could be an important avenue to overcome marketing problems. In fact, contract farming comes with a secured market for the sale of products, a pre-determined price, technical information and inputs supply. The absence of such arrangements is an important constraint for the diversification of crop area. Overall, it emerges that lack of proper institutional and marketing arrangements are responsible for low crop diversification in the state of Bihar.

**Conclusions and Policy Recommendations**

Our analysis shows that despite a thrust on boosting agricultural productivity and a stable political regime, agricultural growth has plummeted since 2012-13. We apply a modified growth diagnostics framework to identify the most binding constraints on agricultural growth. Analysis has revealed that poor functioning of agricultural markets and low level of crop diversification are the reasons for slow or lower agricultural growth in Bihar. To relax these binding constraints, we suggest the following policy measures.

There is a need to design incentive mechanisms to attract private investment in the development of agricultural markets including cold storage and warehousing facilities. Government may step in to provide the basic market infrastructure such as road and market yards to attract the investment. Sharing information about market conditions, particularly prevailing prices on a real-time basis, will help farmers make the right decisions about timing and quantity of products to be sold. Farmer producer organisations (FPOs) have potential to create an enabling environment for direct marketing of agricultural produce. However, FPOs require adequate initial financial support and periodic training for their successful operation.

A comprehensive policy on crop diversification is required to provide incentives for farmers to diversify from a low-value cereal-based system to a high-value fruits and vegetable system. The policy may concomitantly encourage private investment in establishing processing infrastructure for grading, sorting, value addition, etc. of fruits and vegetables. Development of the agro and food processing industry at a cluster level where adequate raw materials are available will generate employment and increase the income of farmers, among others. Contract farming provides a secured market with assured prices for agricultural products. This is important particularly for the growing of perishable products such as vegetables. A suitable legislative measure on contract farming is needed along the lines of the Model Contract Farming Act brought out by the Centre. The Act should provide a level playing field for both producers and agro-commercial firms.